\documentclass[9pt, journal]{IEEEtran}
  \usepackage[font={footnotesize}]{caption}
  \usepackage{amssymb}
  \usepackage{pifont}
  \usepackage{subfig}
  \usepackage{graphicx}
  \usepackage{balance}
  \usepackage{mathtools}
  \usepackage[usenames, dvipsnames]{color}
  \usepackage{algorithm2e}
  \usepackage{enumitem}
  \usepackage{algpseudocode}

\begin{document}

   \title{Using ADC and Backend Capacitors to Authenticate for Free: (Virtually) Free from Database, Enrollment, and Excessive Area/Power Overhead}
  
  \author{Ahish Shylendra, Swarup Bhunia, and Amit Ranjan Trivedi}

  \maketitle
  \begin{abstract}
Detection of counterfeit chips has emerged as a crucial concern. Physically-unclonable-function (PUF)-based techniques are widely used for authentication, however, require dedicated hardware and large signature database. In this work, we show intrinsic \& database-free authentication using back-end capacitors. The discussed technique simplify authentication setup and reduce the test cost. We show that an analog-to-digital converter (ADC) can be modified for back-end capacitor-based authentication in addition to its regular functionality; hence, a dedicated authentication module is not necessary. Moreover, since back-end capacitors are quite insensitive to temperature and aging-induced variations than transistors, the discussed technique result in a more reliable authentication than transistor PUF-based authentication. The modifications to conventional ADC incur 3.2\% power overhead and 75\% active-area overhead; however, arguably, the advantages of the discussed intrinsic \& database-free authentication outweigh the overheads. \textcolor{blue}{\textbf{Full version of this article is published at IEEE TVLSI.}} 


  \end{abstract}
  
\vspace{-1.5em}

  \section{Introduction}
  Counterfeit chips are becoming increasingly common and have emerged as a critical challenge for IC design and fabrication industry [1], [2]. The current mainstream approach to detect counterfeit chips is to embed a physically-unclonable-function (PUF) in the chips. A PUF exploits process variability in design process and produces an uncontrolled and unclonable signature. The output of a PUF in each authentic design is stored in a database. Counterfeit chips are detected by tallying their PUF signature against the database [Fig. 1 (a)]. However, PUF-based authentication has presented two critical challenges: (1) A PUF with adequately large signature size requires considerable area, power, and test overhead, presenting challenges for resource-limited platforms and time to market. (2) With exponential growth of electronic devices, maintaining and tallying against a database of PUF signatures is unwieldy and even impractical for certain applications [3]. 
  
  Countering the above limitations of a PUF, an intrinsic and database-free authentication was presented in [4]. Intrinsic authentication discriminates authentic and counterfeit chips by exploiting their design and fabrication discrepancies. In Fig. 1 (b), for intrinsic authentication, a vendor not only ships an IC to a client but also authentication characteristics ($AC$s) and their expected range of output. $AC$s are designed such that there is a clear distinction between authentic and counterfeit chips. For instance, if a counterfeit chip uses an inferior foundry process or package to save cost (maximize profit) then there will be discrepancies in path delays. Likewise, an adversary pirating an IP may not be aware of the exact layout or routing/placement algorithms/parameters of various design components resulting in path delay discrepancies. Hence, path delays can be used as an $AC$ for authentication. In [4], an intrinsic authentication based on the delay of non-critical paths in scan-chains was presented. 

   \begin{figure}[!t]
       \centering
       \subfloat[][]{\includegraphics[width=4.1cm, height=3.5cm]{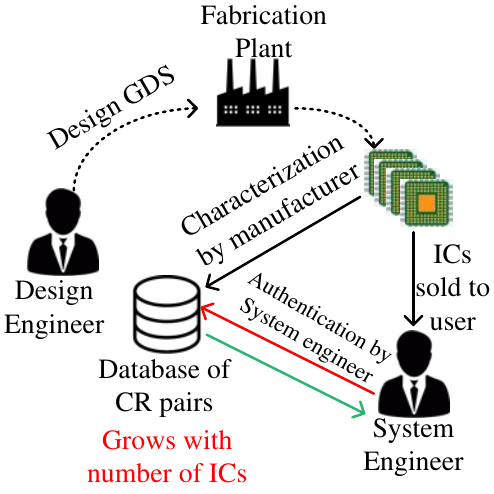}}\quad
       \subfloat[][]{\includegraphics[width=4.1cm, height=3.5cm]{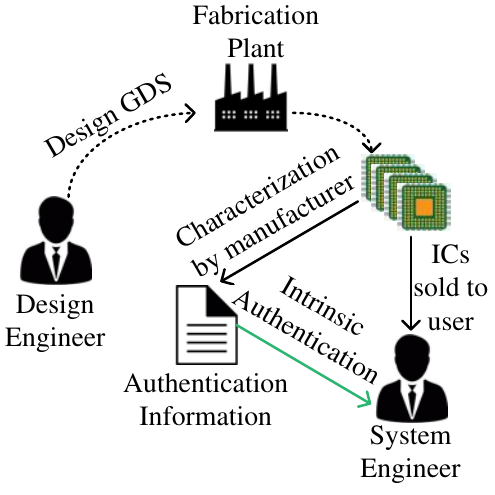}}\\
       \caption{(a) PUF challenge-reponse pair database-based and (b) intrinsic database-free authentication.}
       \label{steady_state}
  \end{figure}

  \begin{figure}[!t]
       \centering
       \subfloat[][]{\includegraphics[width=2.7cm, height=3cm]{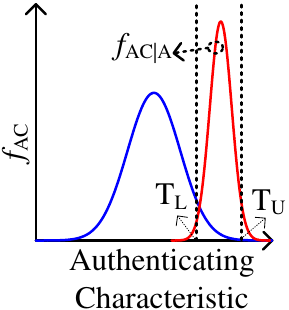}}\quad
       \subfloat[][]{\includegraphics[width=2.7cm, height=3cm]{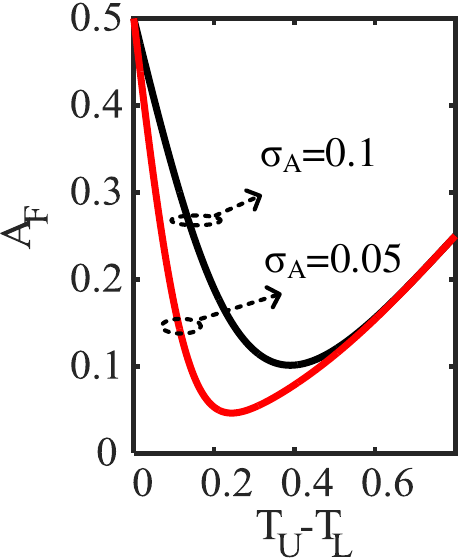}}\quad    
       \subfloat[][]{\includegraphics[width=2.7cm, height=3cm]{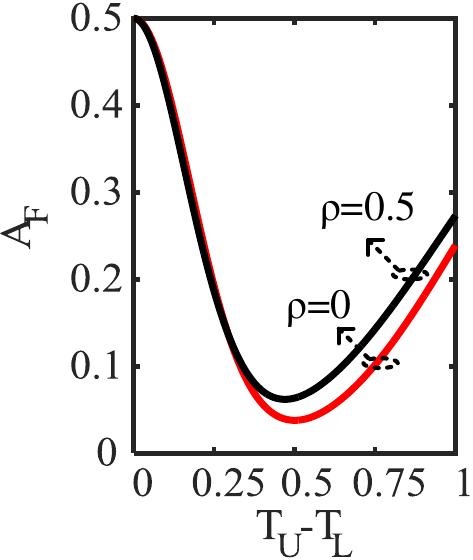}}\quad 
       \caption{(a) PDF of a hypothetical $AC$ in all ICs ($f_{AC}$) and in authentic ICs ($f_{AC|A}$), (b) Authentication failure at varying threshold bounds ($T_U-T_L$) and variance of $f_{AC|A}$. (c) Authentication failure at varying threshold bounds ($T_U-T_L$) and covariance ($\rho$) of ACs.}
       \vspace{-1.5em}
       \label{steady_state}
  \end{figure}
  
  However, $ACs$ in intrinsic authentication should be carefully chosen to minimize false positives and false negatives considering process, temperature, and aging-induced variations. Fig. 2 examines this using probability density function (PDF) for an $AC$ in all (authentic and counterfeit) designs, $f_{AC}$, and in authentic designs, $f_{AC|A}$. For authenticating thresholds $T_L$ $\&$ $T_U$ in Fig. 2, the authentication failure $A_F$ is expressed for counterfeit (C) and authentic (A) chips as
  \begin{subequations}
  \begin{equation}
  A_{F}=P(C)\times P(\tfrac{Pass}{C}) + P(A)\times (1-P(\tfrac{Pass}{A})),  
  \end{equation}
  \begin{equation}
  P(\tfrac{Pass}{C}) = \tfrac{1}{P(C)} \times \big(P(Pass) - P(A) \times P(\tfrac{Pass}{A})\big),
  \end{equation}
  \begin{equation}
  \textstyle P(Pass) = \int_{T_L}^{T_U} f_{AC} dx, \hspace{1mm} P(\tfrac{Pass}{A}) = \int_{T_L}^{T_U} f_{AC|A} dx.
  \end{equation}
  \end{subequations}
  Considering a testcase of $f_{AC} = \mathcal{N}(0.5, 0.1)$ and $f_{AC|A} = \mathcal{N}(0.9, 0.05)$ in Fig. 2(b), the variance in $f_{AC|A}$ limits the minimum $A_F$ under the optimal threshold bound. The observation extends to any $f_{AC}$ \& $f_{AC|A}$. Therefore, $ACs$ showing a small variance from chip-to-chip is necessary for reliable intrinsic authentication. Additionally, multiple $ACs$ can be combined to improve accuracy. For example, an authentication test passes if at least $m$ out of total $n$ $ACs$ pass the test. In this case, authentication failure $A_{F|mult}$ is
  \begin{equation}
  \begin{multlined}
  \textstyle A_{F|mult}=P(C)\times \sum _ {S} P\Big ( \tfrac{Pass}{C, AC_1, AC_2, ... AC_n} \Big ) + \\
  \textstyle P(A)\times \sum _ {S} P\Big ( \tfrac{Fail}{A, AC_1, AC_2, ... AC_n} \Big ). \end{multlined} 
  \end{equation}
  Here, $S$ spans over all possible outcomes of $AC_1...AC_n$. However, $ACs$ must be uncorrelated to minimize $A_{F|mult}$. For example, Fig. 2 (c) considers a test-case with two $ACs$ and accepting a chip if at least one of the two $ACs$ pass the test with $f_{AC} = \mathcal{N}(0.5, 0.25)$, $f_{AC|A} = \mathcal{N}(0.9, 0.1)$ and co-variance $\rho$=0.5. However, in Fig. 2 (c), if $ACs$ are correlated, minimum $A_{F|mult}$ is limited. Considering the above requirements on $ACs$ in intrinsic authentication, this work aims the following contributions:
 
  \begin{figure}[!t]
       \centering
       \includegraphics[width=6cm]{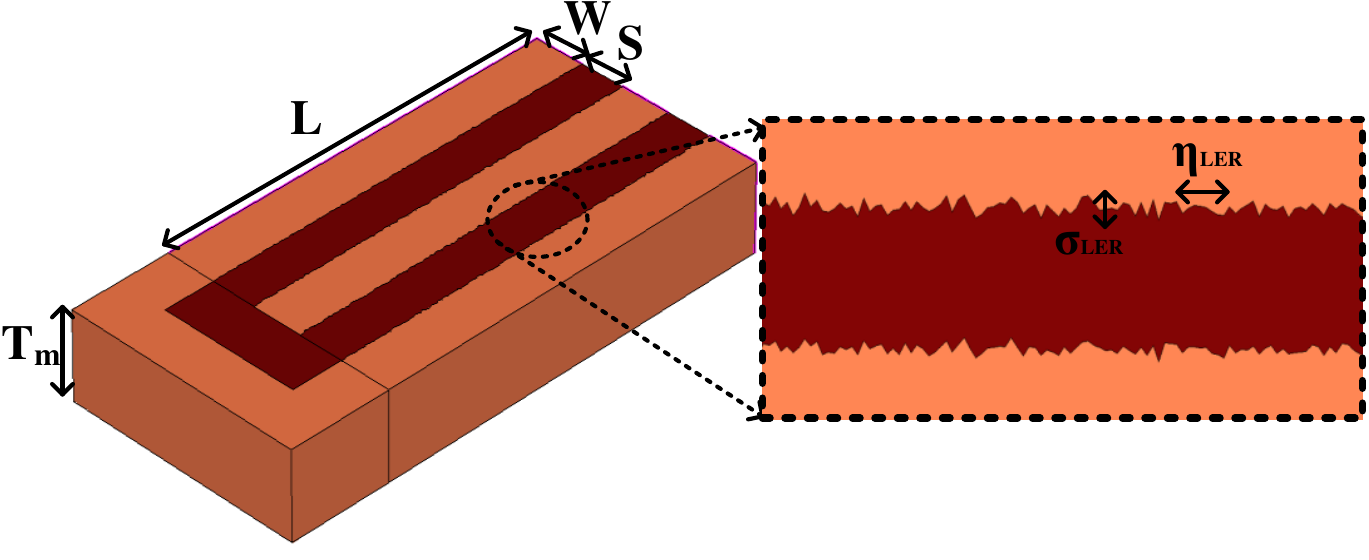}
       \caption{
       Illustration of LER in MOM capacitor.}
       \vspace{-1em}
       \label{steady_state}
  \end{figure}

  \begin{itemize}[leftmargin=*]
    \item \textit{$ACs$ with small variance:} We discuss intrinsic $AC$ based on the standard deviation ($\sigma$) of unit back-end capacitances ($C_u$). Since the back-end capacitances are insensitive to temperature and aging-induced variations, and many unit-sized capacitances are typically used (e.g., in a 10-bit capacitive DAC, 256 $C_u$ are used), $\sigma$ of the unit capacitances can be reliably read for a low variance of $AC$. 

    \item \textit{Low correlation to transistor-based $ACs$:} Since the back-end process steps are typically uncorrelated to the front-end steps, the back-end capacitance-based $ACs$ discussed here have low correlation to transistor-based $AC$ in [4]. Therefore, a combination of $ACs$ presented here with transistor-based $ACs$ can enhance the robustness of authentication.   

    \item \textit{Low overhead $AC$ extraction:} SAR ADC architecture is modified to allow signature extraction using back-end capacitors. Compared to the conventional ADC, modified ADC requires 3.2\% higher power at the same frequency and 75\% higher active-area.
  \end{itemize}

  The present work also extends our prior work [5] where we first discussed the feasibility of intrinsic authentication using back-end capacitors. In this work, we advance the design by considering energy efficient monotonic switching SAR ADC architecture and by avoiding additional parasitics in the signal path to minimize overheads. We also discuss repeated sampling based signature extraction and optimization of ACs for enhanced reliability.

    \begin{figure}[!t]
       \centering
       \subfloat[][]{\includegraphics[width=2.7cm, height=3cm]{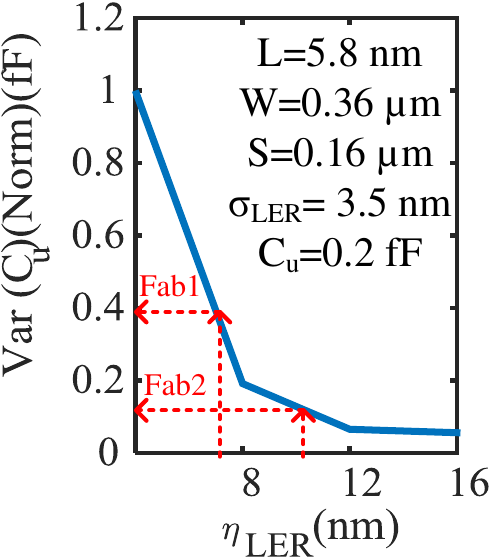}}\quad
       \subfloat[][]{\includegraphics[width=2.7cm, height=3cm]{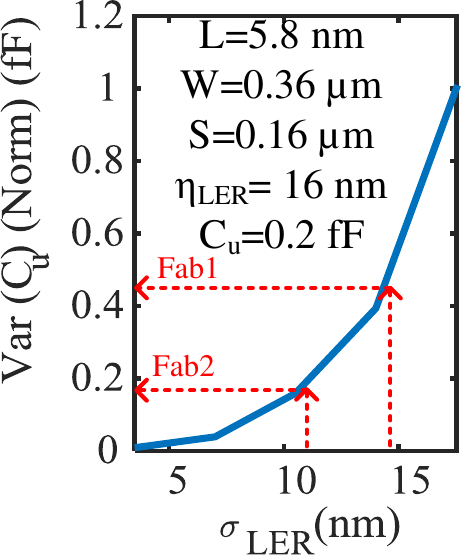}}\quad
       \subfloat[][]{\includegraphics[width=2.6cm, height=3cm]{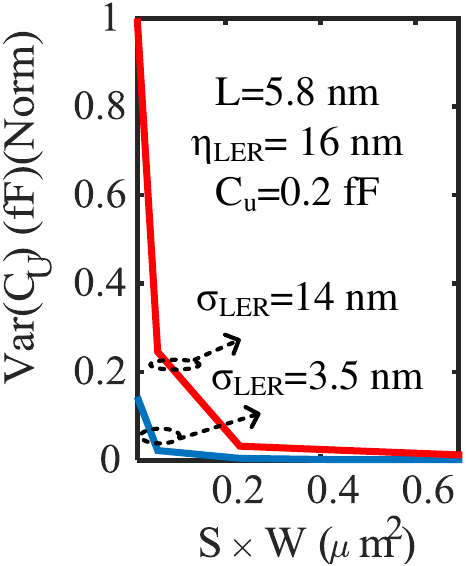}}
       \caption{Variance of MOM capacitor (Normalized) as a function of (a) $\eta_{LER}$, (b) $\sigma_{LER}$, and (c) S (spacing) $\times$ W (metal width) with L=5.8 nm, $\eta_{LER}$=16 nm for different $\sigma_{LER}$.}
       \label{steady_state}
  \end{figure} 
  
  \vspace{-1.2em}

\section{Process-induced Variability in Metal-Oxide\\-Metal (MOM) Capacitors}
  Poly-insulator-poly (PIP), metal-oxide-metal (MOM) and metal-insulator-metal (MIM) are the most widely used back-end capacitors. Currently, MOM capacitors are preferred over the others due to its reduced cost [6] and easy customization for very small capacitances for area/power-constrained designs. Therefore, we focus on MOM capacitor-based $ACs$. Line edge roughness (LER) is the primary source of variability in MOM capacitors [Fig. 3] and originates due to inaccuracies in lithography process such as metal edge roughness, limited resolution, random diffusion of acids, and intrinsic roughness of the resist used. Therefore, LER can be quite foundry-specific depending on the process steps, equipment, and composition of the process materials used in a facility. LER can be characterized by standard deviation ($\sigma_{LER}$) conveying absolute roughness amplitude information and correlation length ($\eta_{LER}$) conveying proximity of the adjacent edges. We consider LER-induced variability in MOM capacitance using Synopsys Sentaurus [7] based TCAD simulations on a hundred capacitor structures. Fig. 4 illustrates the impact of $\eta_{LER}$, $\sigma_{LER}$, and sizing on the variance characteristics of MOM capacitors. From Fig. 4 (a), it can be observed that decreasing the proximity of adjacent edges results in increased variance. In Fig. 4 (b), increasing $\sigma_{LER}$ results in increased variance. Fig. 4 (c) shows MOM capacitor variance at varying capacitor dimensions. Larger area capacitors have smaller variance, in agreement with the Pelgrom's law [6]. In Fig. 4, if a cloned design (Fab 2) is fabricated using different $\eta_{LER}$ and/or $\sigma_{LER}$ or is unaware of the exact capacitor dimensions in the original (Fab 1), discrepancies arise in the capacitance variance.
  
   \vspace{-0.7em}

  \section{ADC-based Intrinsic Authentication exploiting MOM Capacitor Variability }

  \begin{figure}[!t]
       \centering
       \subfloat[][]{\includegraphics[width=8.5cm, height=3.5cm]{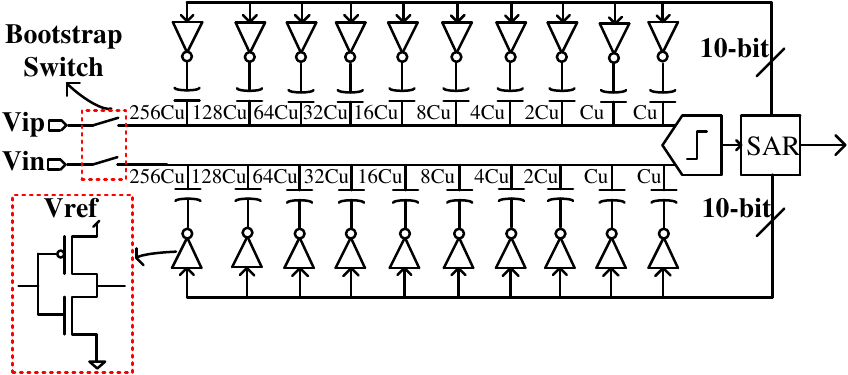}}\\
       \subfloat[][]{\includegraphics[width=6cm, height=3.5cm]{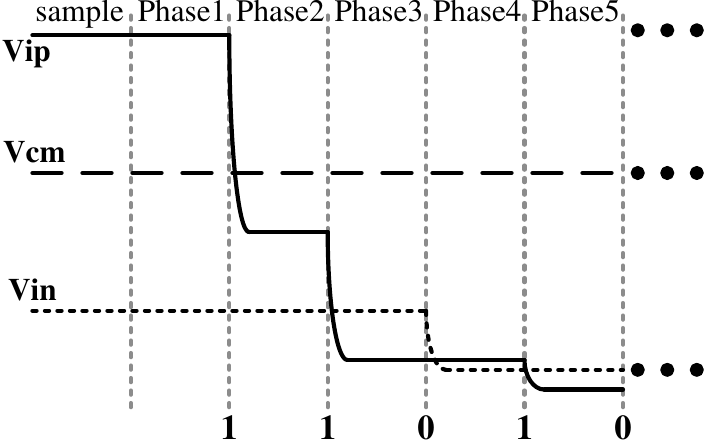}}\\
       \caption{(a) SAR ADC architecture employing monotonic switching, (b) Illustration Monotonic switching procedure.}
       \label{steady_state}
  \end{figure}
  
  \vspace{-0.5em}

  \subsection{Brief Overview of SAR ADC Architecture and Operation}
  Conventional SAR ADC includes charge-redistribution capacitive digital-to-analog converter (CDAC), comparator, and successive approximation logic. SAR ADC with monotonic switching to achieve reduced switching energy has been proposed in [8], [9]. Fig. 5 (a) shows the diagram of 10-bit SAR ADC architecture, where faster reference settling is achieved by using downward switching. Fig. 5 (b) depicts the switching procedure. The input signal is sampled onto the top plates of the capacitors while the bottom plates are connected to V$_{ref}$. Comparator performs the first comparison after the sampling phase. Depending on the comparator output, the bottom plate of the largest capacitor i.e, 256$\times$Cu on V$_{IP}$ terminal is connected to the ground if V$_{IP}$$>$V$_{IN}$, while the bottom plate of the largest capacitor on  V$_{IN}$ side is connected to V$_{ref}$. The connections are reversed on the two sides if V$_{IN}$$>$V$_{IP}$. Above procedure is continued till the decision for the lowest significant bit (LSB) is obtained. 
  \vspace{-1em}
  \begin{figure}[!t]
       \centering      
       \includegraphics[width=9cm, height=3.5cm]{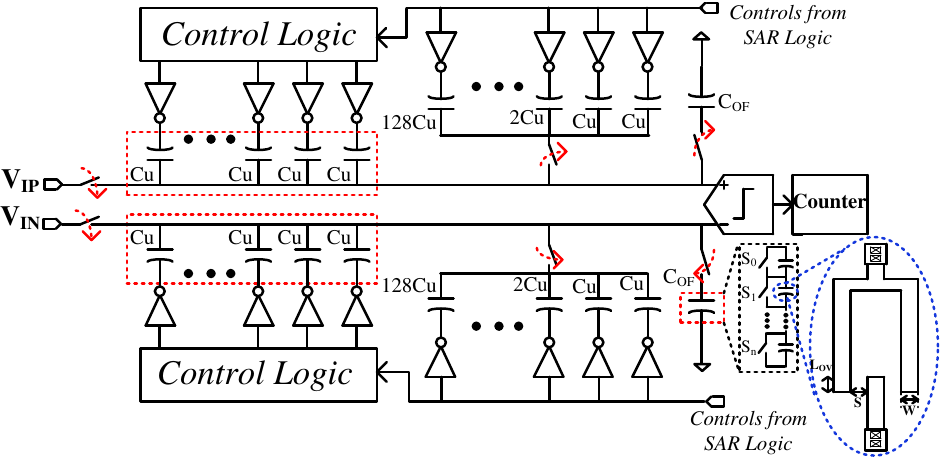}
       \caption{ Modified SAR ADC architecture with authentication capability, Inset: Realizing C$_{OF}$ by lower overlap in metal lines.}
       \label{steady_state}
  \end{figure}
  
    \vspace{-0.5em}
  \subsection{Authentication Scheme}
  \subsubsection{Modified ADC architecture}
  Fig. 6 shows the modified ADC architecture for intrinsic authentication using the back-end capacitors. Our technique uses the fact that larger capacitances in a CDAC are implemented using parallel combination of unit capacitances (C$_u$). Such implementation reduces the overall mismatch of the larger capacitances, but the mismatch in the unit elements cannot be decreased or eliminated. The circuit modifications provide access to each unit element forming a larger capacitor. Additionally, a programmable offset capacitance (C$_{OF}$) with C$_{OF}$ $\sim$ $\sigma_{Cu}$ ($\sigma_{Cu}$ is the standard deviation of C$_u$) is implemented at both ends of comparator. C$_{OF}$ is realized by connecting a programmable number of capacitances in series in Fig. 6 where each small capacitance is formed using lower overlap length.


  \subsubsection{Authentication characteristic extraction}
$AC$ in our scheme is defined by the number $N_{AC}$ of unit capacitance pairs at the positive and negative terminals of the comparator (i.e., $C_{u,P,i}$ \& $C_{u,N,i}$) where $C_{u,P,i} > C_{u,N,i} + C_{OF}$ or $C_{u,N,i} > C_{u,P,i} + C_{OF}$. $N_{AC}$ characterizes the variance of C$_u$. Presently, only the unit capacitors for the maximum significant bit (MSB) in CDAC are utilized for $AC$. Signature extraction involves three steps: In \textit{Step 1}, offset capacitance C$_{OF}$ is connected to the negative input of the comparator in Fig. 7 (i). In \textit{Step 2}, the top and bottom  plates of all C$_u$ in both MSB arrays and C$_{OF}$ are discharged to the ground [Fig. 7 (ii)]. In \textit{Step 3}, one capacitor from each MSB array and C$_{OF}$ are activated by connecting their bottom plates to `1' [Fig. 7 (iii)]. The above procedure is repeated for all C$_u$ and the counter's count is incremented if the comparator's output is one. After this, C$_{OF}$ is activated at the positive input of the comparator. The above steps are repeated and counter's count is incremented when comparator's output is zero. The entire procedure is repeated at varying C$_{OF}$ to obtain the entire $AC$, i.e., $N_{AC}(C_{OF})$. At the foundry, the above experiment is repeated with a significant number of ICs to obtain multiple traces of $N_{AC}(C_{OF})$. From the obtained traces, average trace $N_{AC,AVG}(C_{OF})$ and expected bounds are determined and shipped to a client for intrinsic authentication. 
  
   \begin{figure}[!t]
       \centering      
       \includegraphics[width=8cm, height=5cm]{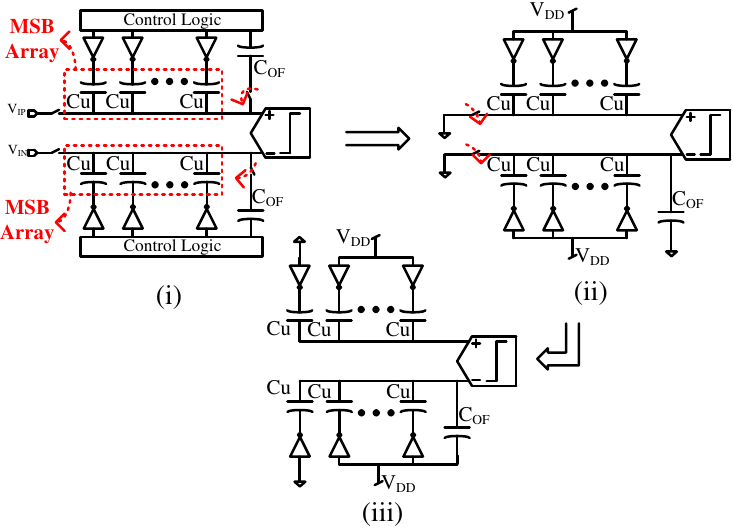}
       \caption{Circuit connections during signature extraction.}
       \label{steady_state}
  \end{figure}


  \vspace{-1.5em}
  \subsection{Simulation Results}
  Circuits utilized to perform the simulation study are implemented using 45nm technology [10]. MOM capacitor model based on [11] is considered. Fig. 8 shows the circuit diagram of dynamic comparator with PMOS input transistors. Outputs Outn and Outp are reset to V$_{DD}$ when clk=0. Differential input pair compares the voltages at the two input terminals when clk=1 and positive feedback action causes the drain potential at either M1 or M2 to become high. Inverters are used at the output stage for rail-to-rail output swing. Offset voltage in a comparator arises due to process variations and is deterministic in nature. Offset voltage can be reduced by using dynamic offset cancellation techniques such as chopping and auto-zeroing [12]. On the contrary, comparator input referred noise is indeterministic in nature and is considered to have a Gaussian distribution profile with zero mean and a variance of 250 nV$^2$ in our simulations [13].  


   \begin{figure}[!t]
       \centering
       \includegraphics[width=6cm]{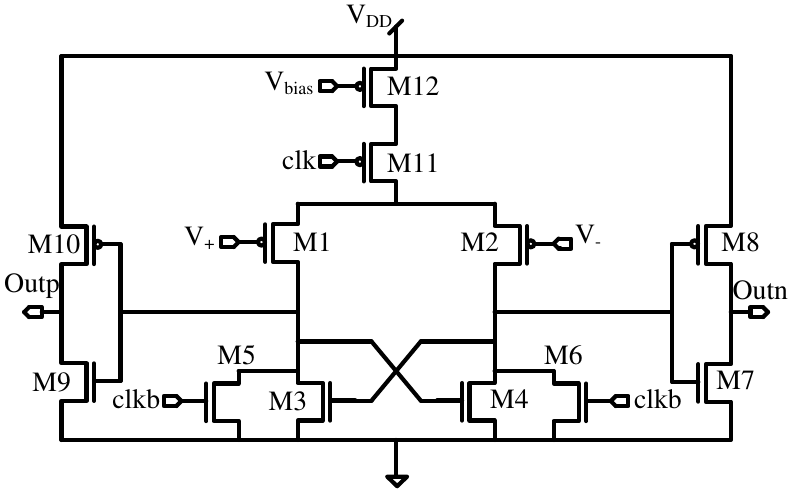}\\
       \caption{Comparator circuit diagram.}
       \label{steady_state}
  \end{figure}

   \begin{figure}[!t]
       \centering
       \subfloat[][]{\includegraphics[width=2.7cm, height=3.1cm]{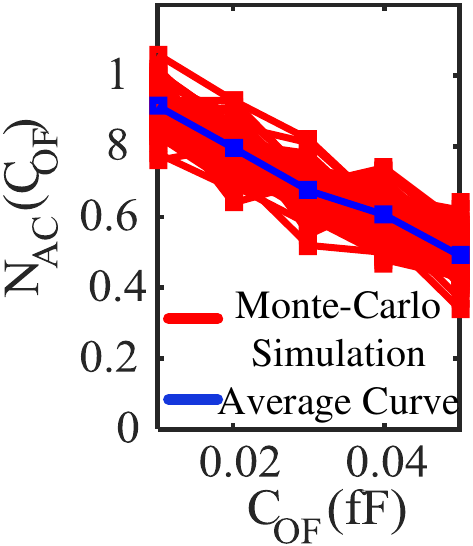}}\quad
       \subfloat[][]{\includegraphics[width=2.7cm, height=3cm]{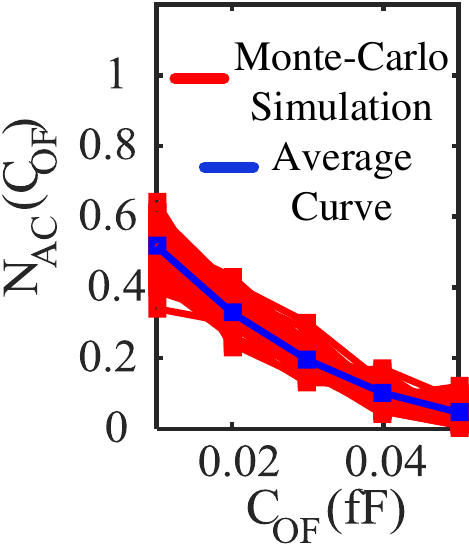}}\quad
       \subfloat[][]{\includegraphics[width=2.7cm, height=3cm]{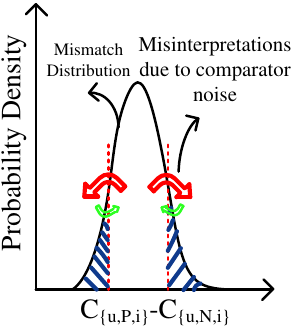}}
       \caption{Quantitative analysis with (a) No repeated sampling, (b) Repeated sampling, and (c) Impact on comparator input noise on mismatch characterization.}
       \label{steady_state}
  \end{figure}
  

Fig. 9 shows the simulation results for $N_{AC}(C_{OF})$ normalized against the total C$_u$ pairs. Monte-Carlo simulations are performed on a hundred designs. The impact of input referred noise is suppressed by repeated comparisons for the considered capacitor pair and taking the dominant comparator output as the final output. Fig. 9 (a) shows the extracted distribution when repeated sampling is not used. In Fig. 9 (a), $AC$ bounds are wider since the input referred noise dominates. Tighter bounds are achieved in Fig. 9 (b) by repeated sampling. Fig. 9 (c) explains the shift in the average trace of $N_{AC}(C_{OF})$ ($N_{AC,AVG}(C_{OF})$) under repeated sampling in Fig. 9(b). In a C$_u$ pair ($C_{u,P,i}$, $C_{u,N,i}$), the distribution of $C_{u,P,i} - C_{u,N,i}$ follows a Gaussian distribution. The comparator identifies when $|$$C_{u,P,i} - C_{u,N,i}$$|$ is more than C$_{OF}$. Since the density of $C_{u,P,i} - C_{u,N,i}$ decreases away from zero, a higher fraction of capacitor pairs is falsely characterized to have a higher mismatch than C$_{OF}$ than the fraction falsely characterized to have a lower mismatch due to comparator noise. Repeated measuring corrects this inaccuracy which also results in $N_{AC,AVG}(C_{OF})$ moving to a lower value. To perform authentication, $N_{AC,AVG}(C_{OF})$ and the expected output bounds are stored. Authenticity of an IC is determined by considering the Euclidean distance ($D_{Auth}$) between its $N_{AC}(C_{OF})$ and $N_{AC,AVG}(C_{OF})$ and by comparing against the bounds. Notably, unlike PUF, where the authenticating database grows with the number of devices, the discussed scheme is database-free where $AC$ test is the same for all devices.


  \vspace{-1.3em}
  \section{Discussion}

  In this section, we discuss the optimization considerations to $AC$ for the optimal number of capacitors to be utilized. Impact of temperature and aging-induced variations on proposed technique is investigated. Subsequently, resource overhead incurred due to modifying conventional SAR ADC architecture has been discussed.

\vspace{-1.2em}

  \subsection{Optimizing Authenticating Characteristics}
We optimize the discussed $AC$ for the following objectives. $\mathcal{O}_1$: Maximize the sensitivity of the authentication metric $D_{Auth}$ to the variance of $C_u$. $\mathcal{O}_2$: Minimize the variance of $D_{Auth}$ (note that $D_{Auth} = ||N_{AC}(C_{OF})-N_{AC,AVG}(C_{OF})||$). At a C$_{OF}$, $N_{AC}$ in Fig. 9 is incremented if the following inequalities hold
  \begin{equation}
  \frac{(C_{u,P,i}+C_{OF})\times V_{REF}}{\sum_{j=1}^N C_{u,P,i} + C_{OF}} + V_{n}(t) < \frac{C_{u,N,i}\times V_{REF}}{\sum_{j=1}^N C_{u,N,i}},
  \end{equation}
  \begin{equation}
  \frac{(C_{u,N,i}+C_{OF})\times V_{REF}}{\sum_{j=1}^N C_{u,N,i} + C_{OF}} + V_{n}(t) < \frac{C_{u,P,i}\times V_{REF}}{\sum_{j=1}^N C_{u,P,i}}.
  \end{equation}
Here, $C_{u,P,i}$ \& $C_{u,N,i}$ is the $i^{th}$ capacitance pair at the positive and negative terminals of the comparator, respectively. $V_{REF}$ is the reference potential to charge the capacitance at the state `1'. $V_{n}$ is time dependent input referred noise to the comparator. $N$ is the number of capacitors used for $AC$. Various quantities in (3) \& (4) follow the distributions below
  \begin{equation}
  C_{u,P,i} \sim C_{u,N,i} \sim \mathcal{N}(C_u,\,\sigma_{cu}^{2}),
  \end{equation}
  \begin{equation}
  V_{n} \sim \mathcal{N}(0,\,\sigma_{n}^{2}),
  \end{equation}
  \begin{equation}
 \textstyle \sum_{j=1}^N C_{u,P,i} \sim \sum_{j=1}^N C_{u,N,i} \sim \mathcal{N}(N\times C_u,\,\sigma_{cu}^{2}/N).
  \end{equation}
Here, $\sigma_{Cu}$ is the standard deviation of $C_u$ mismatch and $\sigma_{n}$ is the standard deviation of input referred noise. From (3-4), although using a fewer number of capacitors (small $N$) is statistically unreasonable, using a very large number of capacitors also reduces the sensitivity of the comparator's output to the mismatch in $C_{u,P,i}$ \& $C_{u,N,i}$ and the output is dominated by $V_n(t)$. Considering the probability distributions in (5-7), Fig. 10 shows variance and mean of $D_{Auth}$ as a function of $N$. In Fig. 10 (a), variance of $D_{Auth}$ reduces at higher $N$ and tighter bounds can be implemented for authentication. However, in Fig. 10(b), sensitivity of $N_{AC, AVG}$ to $\Delta C_{OF}$ also decreases as we increase $N$ (shown mean($D_{Auth}$) traces at $C_{OF}$ = $C_u$/50 $\&$ $C_u$/100), imposing an upper limit on $N$. Trade-offs between Fig. 10 (a) and (b) determine the optimal value of capacitors ($N_{OPT}$) used for authentication. 


   \begin{figure}[!t]
      \centering
       \subfloat[][]{\includegraphics[width=4cm, height=3cm]{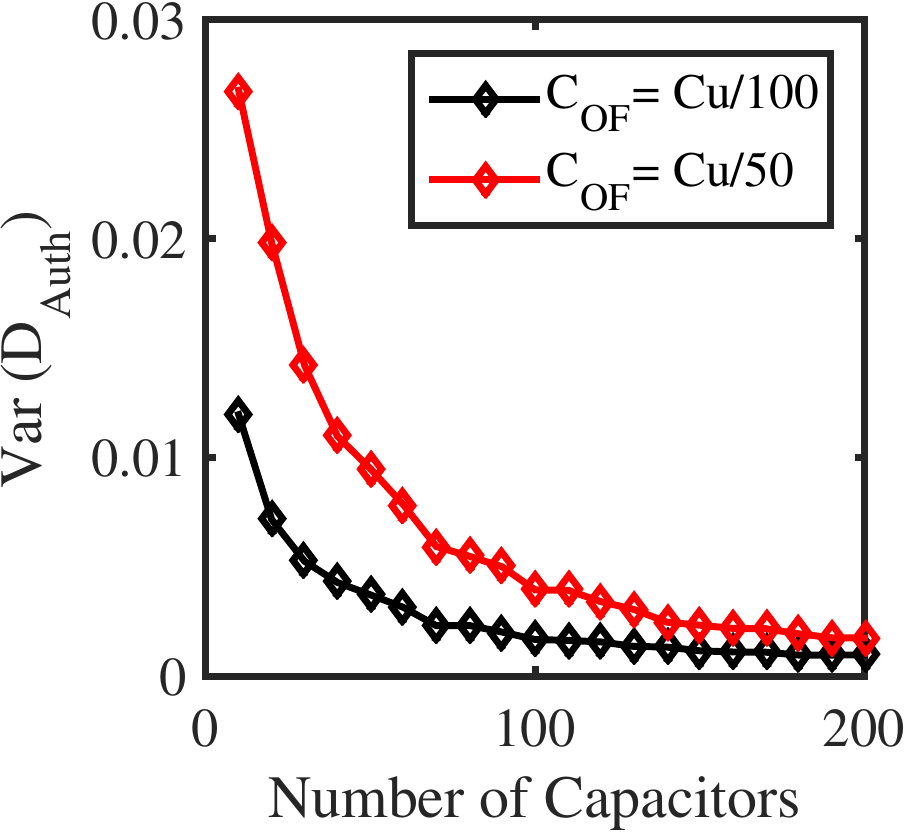}}\quad
       \subfloat[][]{\includegraphics[width=4cm, height=3cm]{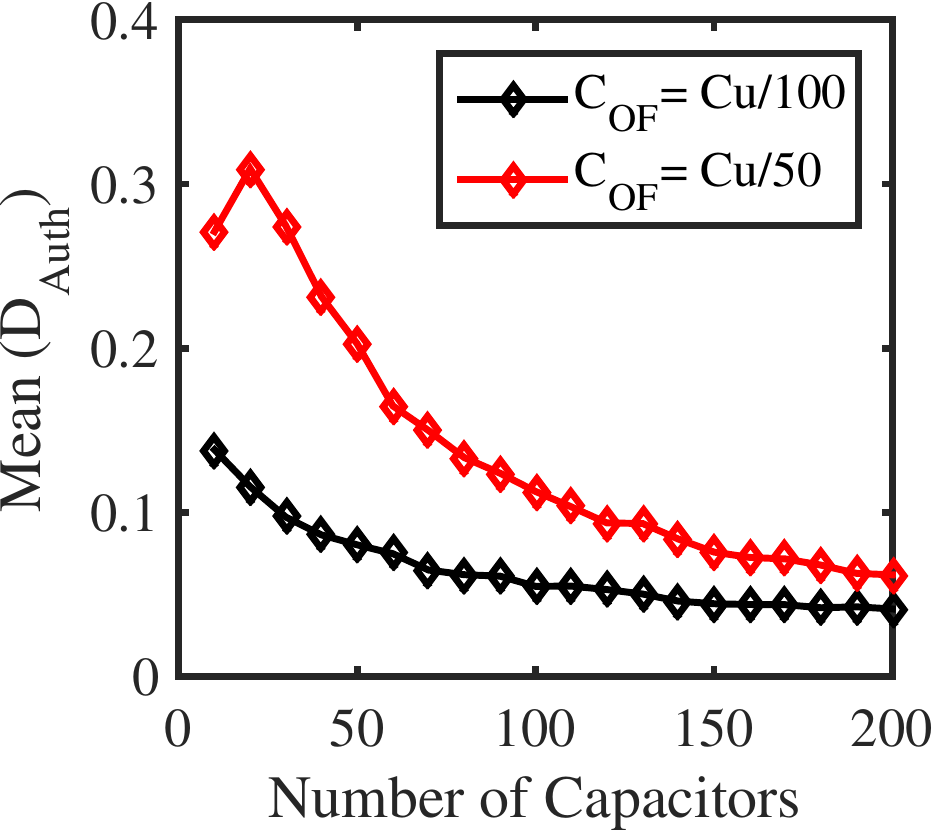}}\\
       \caption{(a) Var($D_{Auth}$) $vs$ Number of Capacitors (N) for different $C_{OF}$, (b) Mean($D_{Auth}$) $vs$  Number of Capacitors (N) for different $C_{OF}$}
       \label{steady_state}
  \end{figure}

   \begin{figure}[!t]
       \centering
       \subfloat[][]{
       \includegraphics[width=4cm, height=3cm]{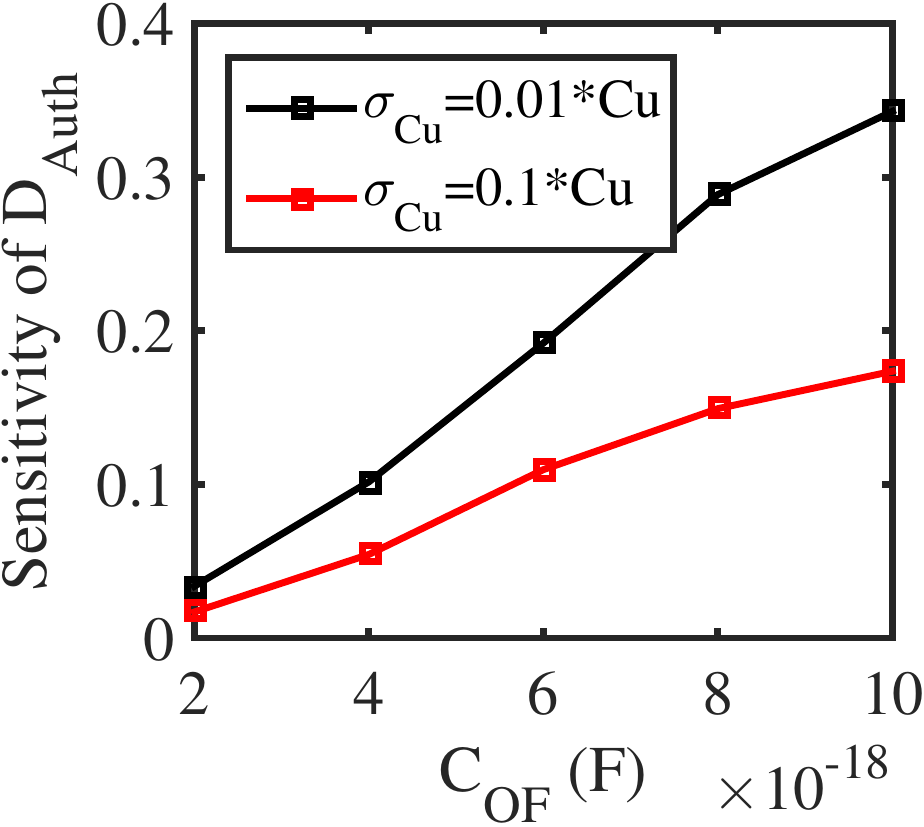}
      }\quad
       \subfloat[][]{\includegraphics[width=4cm, height=3cm]{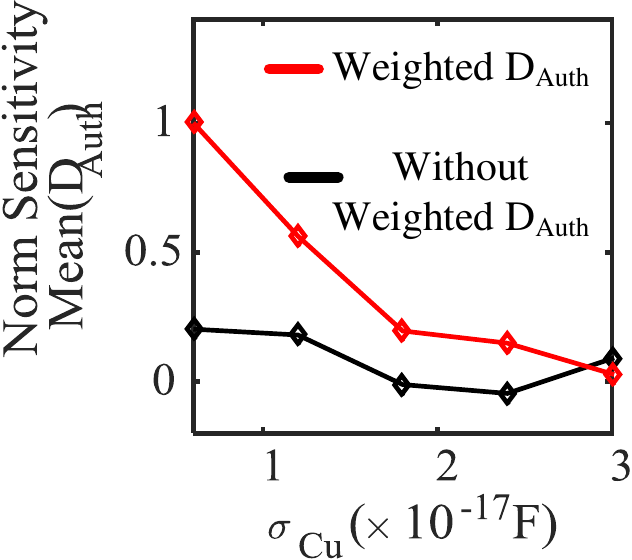}}\\
       \caption{(a) Sensitivity of Mean($D_{Auth}$) vs $C_{OF}$ with variations in $\sigma_{Cu}$, (b) Enhancing Sensitivity of Mean($D_{Auth}$) to variations in $\sigma_{Cu}$ }
       \label{steady_state}
  \end{figure}

The optimization for $\mathcal{O}_1$ can be further enhanced by using a weighted distance of $N_{AC}$ against $N_{AC,AVG}$ at varying C$_{OF}$,      
    \begin{equation}
  D_{Auth, W}= \sqrt{\sum_i \eta_i \times (N_{AC}(C_{OF,i})-N_{AC,AVG}(C_{OF,i}))^2} 
  \end{equation}
where $D_{Auth, W}$ denotes the weighted distance between $N_{AC}$ and $N_{AC,AVG}$, and $\eta_i$ denotes the weight at $C_{OF,i}$. We assign $\eta_i$ proportional to the sensitivity of $D_{Auth}$ to $C_{OF}$ (sensitivity = $\Delta D_{Auth}/\Delta C_{OF}$). In Fig. 11 (a), $D_{Auth}$ is more sensitive to $\sigma_{C_u}$ at higher $C_{OF}$. In Fig. 11(b), the reliability of the authentication scheme is enhanced by performing such weighted analysis, where $D_{Auth, W}$ exhibits higher sensitivity to $\sigma_{C_u}$ variations than $D_{Auth}$ and improves the discernibility even at lower $\sigma_{C_u}$. 

  
   \begin{figure*}[!ht]
       \centering
       \subfloat[][]{\includegraphics[width=4cm, height=4.5cm]{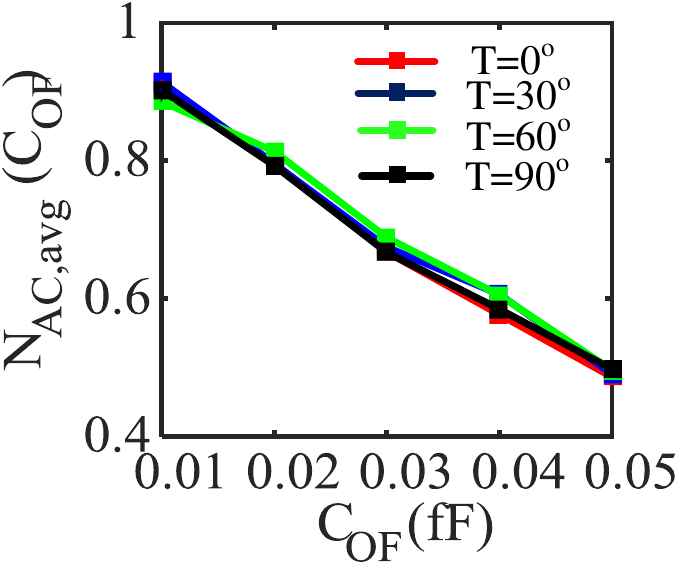}}\quad
       \subfloat[][]{\includegraphics[width=4cm, height=4.5cm]{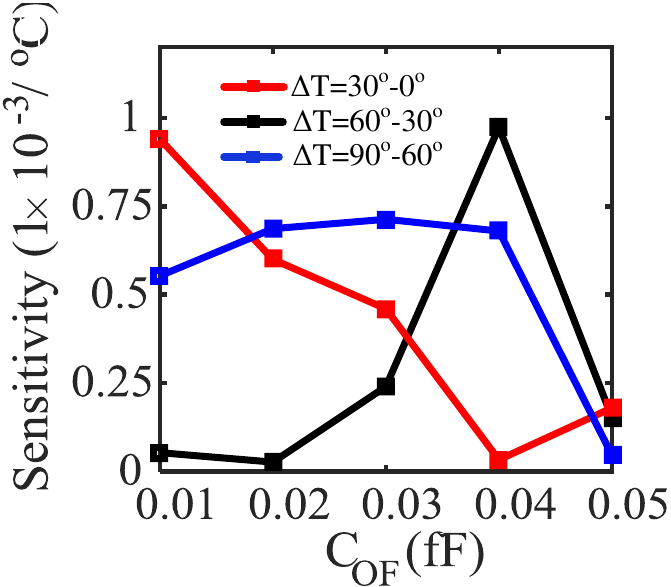}}\quad
       \subfloat[][]{\includegraphics[width=4cm, height=4.5cm]{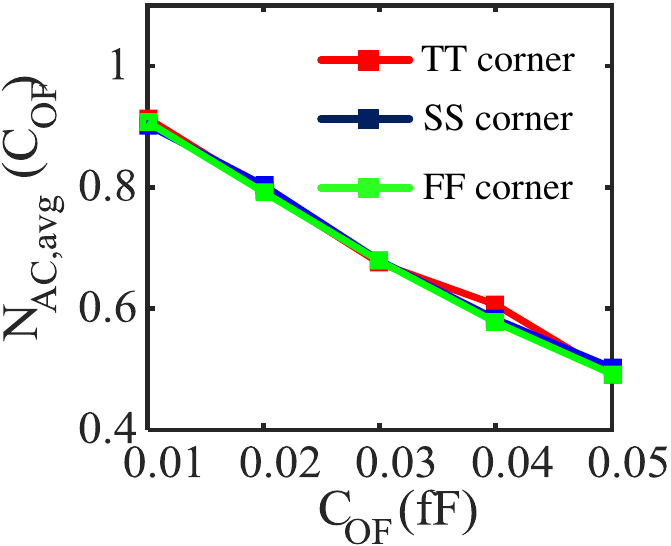}}\quad
       \subfloat[][]{\includegraphics[width=4cm, height=4.5cm]{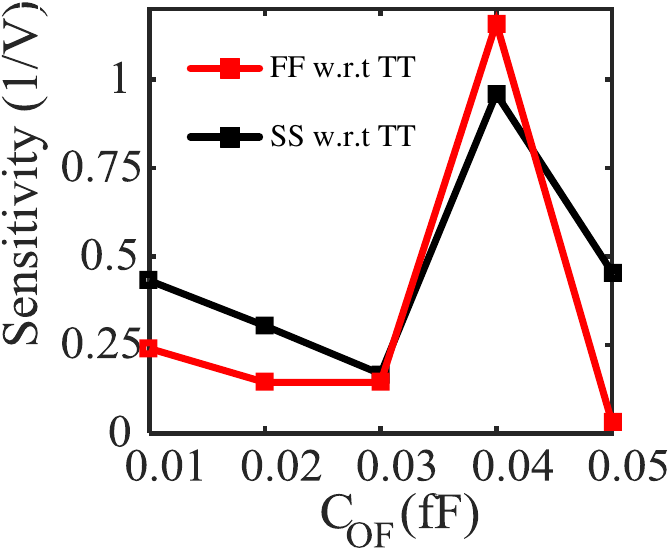}}\quad
        \subfloat[][]{\includegraphics[width=4cm, height=4.5cm]{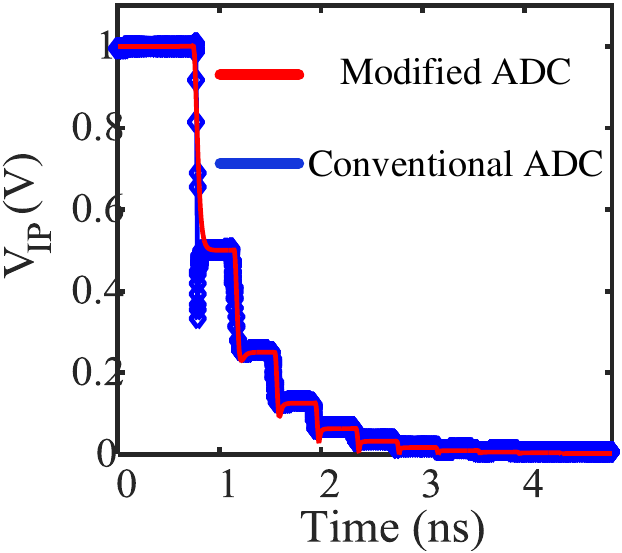}}\quad
       \subfloat[][]{\includegraphics[width=4cm, height=4.5cm]{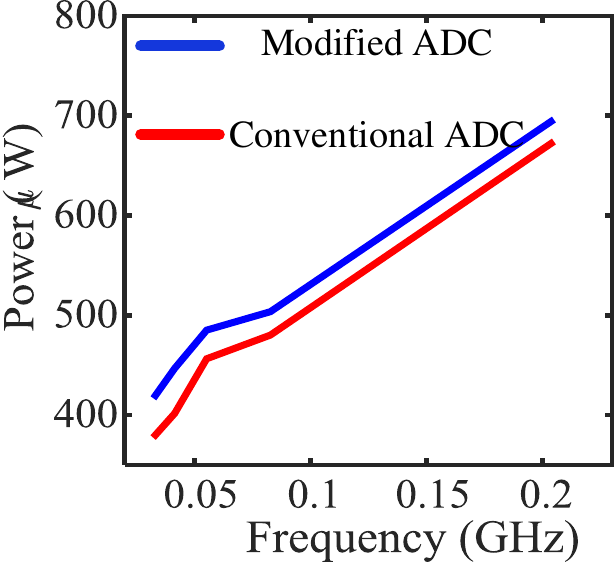}}\\
       
       \caption{(a) Average Distribution $vs$ C$_{OF}$ for different Temperature, (b) Sensitivity Factor $vs$ C$_{OF}$ for different Temperature, (c) Average Distribution $vs$ C$_{OF}$ for different process corners, (d) Sensitivity Factor $vs$ C$_{OF}$ for different process corners, (e) Switching sequence of Conventional and Modified ADC, and (f) Power $vs$ performance analysis..}
       \label{steady_state}
  \end{figure*}
  
  \vspace{-1em}
 
\subsection{Impact of Temperature and Process Variation}
Discussed authentication scheme exploits process variations in capacitors. However, variation of capacitance as a function of temperature is undesired. The expression for capacitance as a function of temperature is given below  
    \begin{equation}
 C(T)=C(T_{0}) \times(1+TC\times(T-T_{0})),
  \end{equation}
where $TC$ denotes the temperature co-efficient for capacitance variation. Note that the temperature sensitivity of the back-end capacitors is much less than in transistors. Additionally from (3-4), the temperature dependence in our scheme is suppressed since the signature extraction depends on the capacitance ratio which is even less sensitive to temperature. In this context, we have studied the impact of temperature variations on $N_{AC, AVG}$ ($C_{OF}$). In Fig. 12 (a), $N_{AC, AVG}$ ($C_{OF}$) is fairly insensitive to temperature variations. Also, sensitivity analysis in Fig. 12 (b) (Temperature Sensitivity = $\Delta N_{AC, AVG} (C_{OF})/\Delta T$) shows that capacitance ratio based signature extraction is highly resilient to temperature variations, providing reliable estimate of $N_{AC, AVG} (C_{OF})$.

Moreover, process variability in CMOS circuitry only has a second order effect to the robustness of discussed authentication since the authentication primarily depends on capacitors. We earlier discussed comparator offset reduction using [12]. In Fig. 12 (c), we have studied the impact of process corner variation in transistors of peripheral circuitry on $N_{AC, AVG}$ ($C_{OF}$). Additionally, sensitivity analysis of the discussed scheme for different process corners computed as $\Delta N_{AC, AVG} (C_{OF})/\Delta V_{th}$ is shown in Fig. 12 (d). Global $V_{th}$ variations in transistors (especially in those used in comparator) does not affect extracted signature $N_{AC, AVG}$ ($C_{OF}$) significantly and efficiently captures mismatch information from capacitors. 
  \vspace{-1.2em}

  \subsection{Overheads to the Regular Operation of ADC}
  
  Resource overhead in the modified SAR ADC arises as a result of using additional inverters to access unit elements of MSB capacitor. Further, control logic used for controlling signature extraction and ADC operation leads to additional overhead.
 Fig. 12 (e) depicts the HSPICE simulated switching sequence of conventional \&  modified ADCs. Incorporated modifications preserve the switching sequence. 
The conventional and modified ADC architectures can be reliably operated at a maximum frequency of 205.2 MHz, thus, modifications incorporated does not affect the maximum operating frequency. Fig. 12 (f) shows the plot for power $vs$ performance analysis where modified ADC consumes $\sim$ 3.2\% higher power than the conventional design while operating at 205.2 MHz. Further, modified ADC has an active area overhead of 75\% when compared to conventional ADC. Active area is computed by adding up $L\times W$ of all the transistors. It should be noted that overhead can be further reduced by decreasing $N$; leading to simplified control logic design as well as reducing number of additional inverters. However, this can critically degrade authentication reliability.


  \vspace{-1.4em}
  
   \subsection{Comparison between PUF and Intrinsic Authentication}
   
\begin{table}[t]
\caption{Comparison between PUF based authentication and our scheme.}
\scalebox{0.8}{
\begin{tabular}{|c|c|c|c|}
\hline
Particulars              & RO-PUF                                                                                                                 & Arbiter-PUF                                                                                                          & This work                                         \\ \hline
Temperature Sensitivity  & 42.25\%                                                                                                                & 68.83\%                                                                                                              & 5.11\%                                            \\ \hline
Aging induced Variations & 87.97\%                                                                                                                & 66.34\%                                                                                                              & 3\%                                               \\ \hline
Average Power (W)        & \begin{tabular}[c]{@{}c@{}}3.83$\times 10^{-6} \times N_{INVs}$\\ per response bit\end{tabular} & \begin{tabular}[c]{@{}c@{}}3.347$\times 10^{-6} \times N_{Ds}$\\ per response bit\end{tabular} & 696.13$\mu$                                       \\ \hline
Database-free            & \textcolor{red}{$X$}                                                                                & \textcolor{red}{$X$}                                                                              & \textcolor{blue}{$\checkmark$} \\ \hline
Dedicated Hardware       & \textcolor{red}{$\checkmark$}                                                                       & \textcolor{red}{$\checkmark$}                                                                     & \textcolor{blue}{$X$}          \\ \hline
\end{tabular}}
\end{table}
   
 Table I summarizes the comparison between PUF based authentication [14, 15] and our scheme. PUF-based authentication requires dedicated hardware and database of challenge-response pairs (CRPs). Whereas our scheme is intrinsic and can utilize the existing components (in this paper, an ADC) for authentication. In Table I, the sensitivity for temperature and aging-induced variations in PUF designs has been computed as percentage change in average frequency difference (RO PUF) and average delay difference (arbiter PUF) for 30$^o$C change in temperature and 10\% change in V$_{th}$ of transistors, respectively. Proposed scheme manifests better resilience to temperature and aging induced variations compared to RO \& arbiter PUFs by relying on the back-end capacitors which are more temperature/aging-insensitive than front-end components. Further, power consumption in PUF scales with CRPs where the number of CRPs depends on the number of on-field devices. Power consumption of PUF in Table I is shown as a function of the number of inverters (N$_{INVs}$) in RO PUF \&  number of delay elements (N$_{Ds}$) in Arbiter PUF decided by N$_{CRPs}$ [14], [15]. Meanwhile, the proposed solution is database-free; hence, the power dissipation of the scheme is not determined by the number of on-field devices. Thus, the proposed scheme can better suit applications with a very large number of on-field devices. 


%
%
%
%
  


  
    \vspace{-2em}

  \section{Conclusion}
Mismatch in back-end capacitors is efficient source for authentication. In this work, MOM capacitor based authentication technique has been proposed using SAR ADC to extract the signature. Proposed authentication scheme does not require dedicated hardware as it utilizes invariably present ADC block. Moreover, our scheme is database-free unlike PUF based authentication. Discussed authentication scheme achieves better reliability than transistor based techniques as back-end capacitors are fairly insensitive to temperature and aging induced variations. Additionally, power consumption in our scheme does not depend on number of on-field devices, which is not the case with PUFs. Although, incorporated modifications to the conventional SAR ADC architecture incur 3.2\% power overhead and 75\% active area overhead, advantages of our intrinsic and database-free authentication can supersede overhead in many applications.



 \section{Acknowledgment}
This work is an outcome of research undertaken in the project supported by Semiconductor Research Corporation (SRC) under grant 2712.022.


  %
  \balance
  \bibliographystyle{abbrv}
  

\begin{thebibliography}{00}

  \bibitem{1} A. Herkle, J. Becker, and M. Ortmanns, "Exploiting Weak PUFs From Data Converter Nonlinearity—    E.g., A Multibit CT $Delta-Sigma$ Modulator," in IEEE Trans. on Circuits and Systems I: Regular Papers, vol. 63, no. 7, pp. 994-1004, July 2016.

  \bibitem{2} U. Guin, K. Huang, D. DiMase, J. M. Carulli, M. Tehranipoor, and Y. Makris, "Counterfeit Integrated Circuits: A Rising Threat in the Global Semiconductor Supply Chain," in Proc. of the IEEE, vol. 102, no. 8, pp. 1207-1228, Aug. 2014.





  \bibitem{5} Q. Ma, C. Gu, N. Hanley, C. Wang, W. Liu, and M. O'Neill, "A machine learning attack resistant multi-PUF design on FPGA," Asia and South Pacific Design Automation Conference (ASP-DAC), Jeju, Korea (South), 2018, pp. 97-104.


  \bibitem{6} Y. Zheng, X. Wang, and S. Bhunia, "SACCI: Scan-Based Characterization Through Clock Phase Sweep for Counterfeit Chip Detection," in IEEE Trans. on Very Large Scale Integration (VLSI) Systems, vol. 23, no. 5, pp. 831-841, May 2015.

  
   \bibitem{8} A. Shylendra, S. Bhunia, and A. R. Trivedi, "Intrinsic and Database-free Watermarking in ICs by Exploiting Process and Design Dependent Variability in Metal-Oxide-Metal Capacitances," in Proc. of the International Symposium on Low Power Electronics and Design, 2018, pp. 1-6.

  \bibitem{9} H. Omran, H. Alahmadi, and K. N. Salama, "Matching Properties of Femtofarad and Sub-Femtofarad MOM Capacitors," in IEEE Trans. on Circuits and Systems I: Regular Papers, vol. 63, no. 6, pp. 763-772, June 2016.

  \bibitem{10} TCAD Sentaurus Manual, USA, CA, Synopsys, Inc., 2017.

  \bibitem{11} C. C. Liu, S. J. Chang, G. Y. Huang, and Y. Z. Lin, "A 10-bit 50-MS/s SAR ADC With a Monotonic Capacitor Switching Procedure," in IEEE Journal of Solid-State Circuits, vol. 45, no. 4, pp. 731-740, April 2010.


  \bibitem{12} M. Ding, P. Harpe, Y. H. Liu, B. Busze, K. Philips, and H. de Groot, "A 46 $\mu W$ 13 b 6.4 MS/s SAR ADC With Background Mismatch and Offset Calibration," in IEEE Journal of Solid-State Circuits, vol. 52, no. 2, pp. 423-432, Feb. 2017.



  \bibitem{13} https://www.eda.ncsu.edu/wiki/FreePDK
  
  \bibitem{14} M. Ding, P. Harpe, Y. Liu, B. Busze, K. Philips, and H. de Groot, "A 46 $\mu \text{W}$ 13 b 6.4 MS/s SAR ADC With Background Mismatch and Offset Calibration," in IEEE Journal of Solid-State Circuits, vol. 52, no. 2, pp. 423-432, Feb. 2017.
  
  \bibitem{15} F. Maloberti: "Analog Design for CMOS VLSI Systems"; Kluwer Academic Publishers, Dordrecht, 2001.

  \bibitem{16} W. P. Zhang, and X. Tong, "Noise Modeling and Analysis of SAR ADCs," in IEEE Trans. on Very Large Scale Integration (VLSI) Systems, vol. 23, no. 12, pp. 2922-2930, Dec. 2015.
  
  \bibitem{17} Y. Cao, L. Zhang, C. Chang, and S. Chen, "A Low-Power Hybrid RO PUF With Improved Thermal Stability for Lightweight Applications," in IEEE Trans. on Computer-Aided Design of Integrated Circuits and Systems, vol. 34, no. 7, pp. 1143-1147, July 2015.
  
  \bibitem{18} L. Lin, D. Holcomb, D. K. Krishnappa, P. Shabadi, and W. Burleson, "Low-power sub-threshold design of secure physical unclonable functions," in ACM/IEEE International Symposium on Low-Power Electronics and Design (ISLPED), 2010, pp. 43-48.

  \end{thebibliography}
  \end{document}